\documentclass{article}
\usepackage{spconf,amsmath,graphicx,amsthm,amssymb}
\usepackage[T1]{fontenc} 
\usepackage[cp1250]{inputenc}
\usepackage[slovak]{babel} 
\usepackage[backend=biber, style=ieee]{biblatex}
\usepackage{tipa}
\bibliography{strings,refs,icassp}

\def\vecs{{\mathbf s}}
\def\vecw{{\mathbf w}}

\def\vecb{{\mathbf b}}
\def\vecr{{\mathbf r}}

\newtheorem{example}{Example}
\title{Phoneme discrimination using neurons with symmetric nonlinear response over a spectral range}
\name{Ondrej \v Such$^{1,2}$, Ondrej \v Skvarek$^{2}$, Martin Klimo$^{2}$ \thanks{Research partially supported by grants APVV-0219-12 and VEGA 2/0112/11 . Usage of computational facilities of University of \v Zilina and University of Matej Bel is gratefully acknowledged.}}
\address{
${}^{1}$ Mathematical Institute of Slovak Academy of Sciences, \v Dumbierska 1, Bansk\'a Bystrica\\
${}^2$ University of \v Zilina, Univerzitn\'a 8215, \v Zilina, Slovakia
 }
\begin{document}
%\ninept
%
\maketitle
\begin{abstract}
We consider the ability of a very simple feed-forward neural network to discriminate phonemes based on 
just relative power spectrum. The network consists of two neurons with symmetric nonlinear 
response over a spectral range. The output of the neurons is subsequently fed to a comparator. We show that often this is enough to achieve complete separation of data. We compare the performance of found discriminants with that of more general neurons. Our conclusion is that not much is gained in passing to real-valued weights. More likely higher number of neurons and preprocessing of input will yield better discrimination results. The networks considered are directly amenable to hardware (neuromorphic) designs. Other advantages include interpretability, guarantees of performance on unseen data and low Kolmogoroff's complexity.
%The abstract should appear at the top of the left-hand column of text, about
%0.5 inch (12 mm) below the title area and no more than 3.125 inches (80 mm) in
%length.  Leave a 0.5 inch (12 mm) space between the end of the abstract and the
%beginning of the main text.  The abstract should contain about 100 to 150
%words, and should be identical to the abstract text submitted electronically
%along with the paper cover sheet.  All manuscripts must be in English, printed
%in black ink.
\end{abstract}
\section{Introduction}
\label{sec:intro}

Artificial neural networks have emerged as one of the most powerful tools in speech recognition \cite{Hinton12}, and more generally in machine learning \cite[Chapter 5]{BishopMachineLearning}. If there is a downside to employing neural networks, it is their opaqueness. Much like a human brain which inspired them, it is often not clear why they work so well. It is however possible, and even advisable \cite[page 148]{Ackley1985147}, to address this opaqueness by adopting design principles that will provide guarantees about their performance in situations that did not occur during training.

In our work we introduce a new class of neural networks that may be used for discrimination between phonemes. They arose in connection with investigation of computational power of memristor based networks.   
As such, they use predominantly min and max processing primitives. There are other approaches that use the same processing elements \cite{BaduraPhd}, \cite{FoltanPhd}, \cite{Foltan}, \cite{FoltanSmiesko}, \cite{Badura1}, \cite{Badura2} as well as a vast body of research on more general fuzzy logic systems. The difference in our work is that we strived to achieve two invariance properties of the resulting network. 

First, we demand that the output of the neural network is \emph{balanced} i.e. invariant with respect to loudness. We achieve this requirement by restricting our attention to neural networks carrying out computation 
\begin{align}
f(\vecs) = f_1(\vecs) - f_2(\vecs),
\end{align}
where $\vecs$ represents sound and $f_1, f_2$ are nonlinear functions that grow additively with increase in loudness i.e.
\begin{align}
f_1(\vecs) - f_1(\vecs') = f_2(\vecs) - f_2(\vecs') = d,
\end{align}
if sounds $s$ and $s'$ differ purely by $d$ decibels in loudness. 
%This requirement may explain the psychoacoustic phenomenon of Cohen's experiment  reported in \cite{Hermansky11}.

%This requirement is of course an oversimplification of human hearing, since it may lead to detection of % articulations in a noise signal that is well below hearing threshold. 

\begin{figure}[htb]
\includegraphics[width=4.0cm, height=2.5cm]{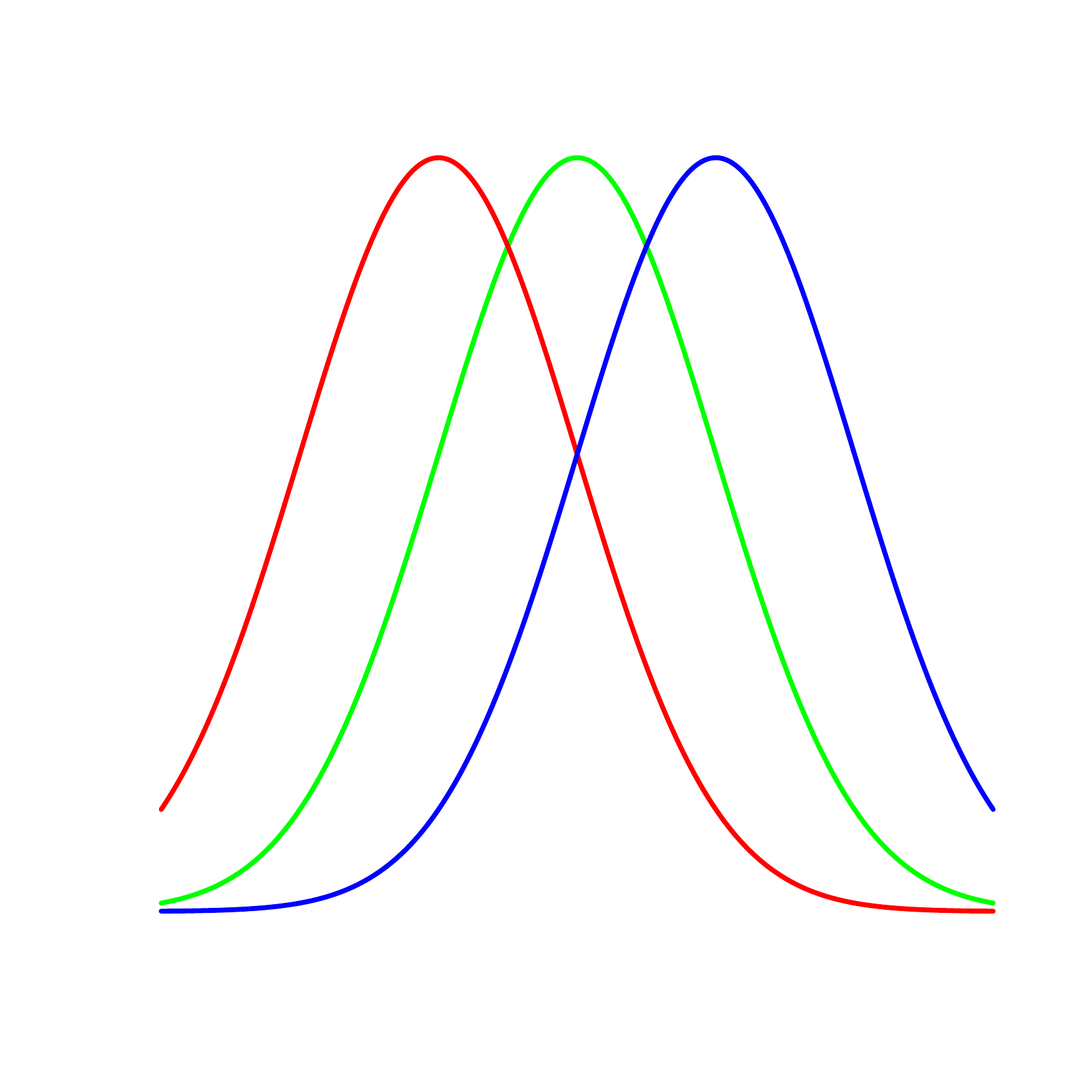}
\begin{centering}
\caption{Sketch of power spectra of three signals with a similar formant frequency.}
\label{fig:norm3}
\end{centering}
\end{figure}

Secondly, we demand a symmetric response of neurons $f_1, f_2$ over a spectral range.  Write $\vecs = (s_1,\ldots, s_n)$ for the discrete log-periodogram of a sound, i.e. $s_i$ represents the log of power at frequency $(i-1)\frac{f_S}{2n}$. A \emph{spectral range} $\vecr$ is any subsequence $ (s_i, s_{i+1}, \ldots, s_j)$ with $i\leq j$. Symmetric response over a spectral range $\vecr$ of a neuron  represented by $k$-ary function $f$   means  that 
\begin{align} 
f(\vecr) = f(\sigma(\vecr))
\end{align}
for any permutation $\sigma$ of $k$ elements.  This requirement is a strong form of requiring that the response be  invariant to small shifts of formant frequencies. Consider a family of signals, spectra of three are sketched in Figure \ref{fig:norm3}. 
% x = seq(-3,3,0.01)
% y1 = dnorm(x, mean = 0)
% y2 = dnorm(x, mean=-1)
% y3 = dnorm(x, mean=1)
% plot(x,y1, ann = FALSE, axes = FALSE, type = "l", col = "green", lwd=3)
% lines(x,y2, ann = FALSE, axes = FALSE, col="red", lwd= 3)
% lines(x,y3, ann = FALSE, axes = FALSE, col="blue", lwd= 3)
Suppose one wants to construct a function with strong response for signals in this family with formants varying between frequencies $\omega_1$ and $\omega_2$. If one restricts oneself to linear forms, there is essentally a single expression that has uniformly strong response for all signals in the family, namely
\begin{align}
s_{i} + s_{i+1} + \cdots + s_{j}.
\end{align}
 %if the ordered discrete spectrum is $\vecs = (s_1, \ldots, s_n)$.
This is not true, if one tries expressions from nonlinear algebras, even as simple as the algebra generated by binary $\min$ and $\max$ functions. For instance, consider functions
\begin{align}
& \max(s_i, s_{i+1}, \ldots, s_j) \\
& \max\nolimits_2(s_i, s_{i+1}, \ldots, s_j)
\end{align}
where $\max_2$ denotes the second largest element of the set. Both of hese functions have strong and symmetric response uniformly over all signals in the family.
%Another way to look at the second requirement is that we impose an analogue of commutativity of summation.  

In our work we shall be concerned with induced \emph{$B$-classifiers} that determine  the class of a phoneme based on comparing $f(\vecs)$ with a threshold $\theta$, and a special subclass of \emph{$Z$-classifiers}, for which $\theta = 0$.

\section{Optimization methods}

A vast majority machine learning techniques such as support vector machines, neural networks, or various regressions have a parameter space that forms a Riemannian manifold. Consequently with these techniques one may use gradient based optimization mechanisms and sometimes even convex optimization. The situation with our class of neural networks is different. We need to find an optimal structure in a discrete parameter space.  There are two hurdles that need to be addressed. 

First, in the absence of a clever trick, one needs to search the discrete space in a reasonable amount of time. One may opt for a local search whereby an initial network is optimized by small twists, or perhaps by a genetic algorithm. The disadvantage of the former is that the search may end in a suboptimal local minimum, whereas the latter may take a long time to find a good network. In our work we opt for exhaustive search of polynomially growing family, namely we will consider only functions of the form
\begin{align}
f(\vecs) = q_1(\vecr_1) - q_2(\vecr_2),
\end{align}
where $q_1,q_2$ are quantiles (e.g. $\max$, $\max_2$), and $\vecr_1, \vecr_2$ are spectral ranges. If one considers $\vecr_i$  of length at most $L$ and the whole spectrum has $N$ power points, then there are $O(N^2 L^4)$ such functions and  an exhaustive search is feasible.

Secondly, one needs to define a goodness criterion. Typically this is classification success. However, in the case when multiple discrimination functions achieve perfect separation on training data, a more refined criterion is needed. In this context one needs to distinguish between training $Z$-classifiers and $B$-classifiers. For $B$-classifiers, usual Fisher discriminant may be used, which is defined as 
\begin{align}
F_B(f) := \frac{(\mu_1 - \mu_2)^2}{\sigma_1^2 + \sigma^2_2},
\end{align}
where $\mu_i, \sigma_i^2$ are means and variances of evaluations of discriminant function $f$ over the two classes. For $Z$-classifiers we have maximized the following secondary tie-breaking criterion
\begin{align}
F_Z(f) := \min\Bigl(\frac{\mu_1^2}{\sigma_1^2}, \frac{\mu_2^2}{\sigma_2^2}\Bigr),\quad \textrm{if $\textrm{sign}(\mu_1)\not=\textrm{sign}(\mu_2)$}.
\end{align}

For our data set we used  discrete spectra created by A. Buja, W. Stueltze and M. Maechler and used in work \cite{HastieBuja}. It is freely available in ElemStatLearn package of R statistics software as well as online \cite{EnglishPhonemes}. The data set contains spectra of five english phonemes computed from TIMIT database.

We have used custom-built C++ software for results obtained in the next two sections,  R for verification, graphing and Nelder-Mead optimization in section 4 and Matlab for computing data in Figure \ref{fig:prechod}.

\section{Results}

\begin{figure*}[!htb]
  \centering
  \centerline{\includegraphics[width=16cm]{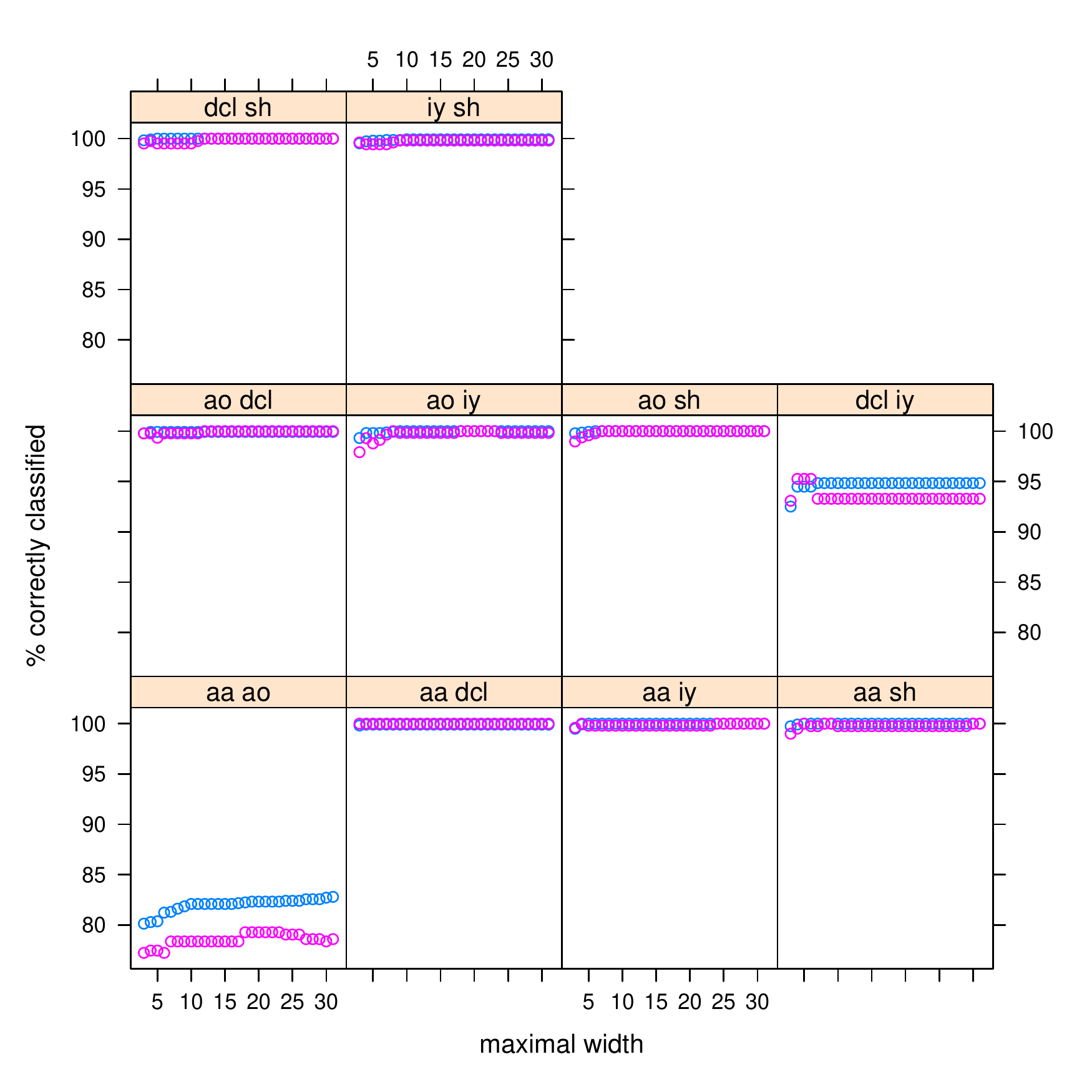}}
  \vskip -0.9cm
  \caption{Train and test success rates for various pairwise $Z$-classifiers}
  \label{fig:results1}
  \end{figure*}

Figure \ref{fig:results1} presents the results of classification by $Z$-classifiers. From the graphs it is clear that in majority of cases, the discriminant functions we considered are able to completely separate the two classes. Moreover, separation occurs with relatively short spectral ranges, with size at most 12. There are just two cases where separation does not occur and that is discrimination of pairs `aa'-'ao' and `dcl'-'iy'.

It is interesting to note that passing from $Z$-classifiers to more general $B$-classifiers does not improve results very much as can be seen in Table \ref{tab:results1}.
\begin{table}[!ht]
\begin{center}
\begin{tabular}{|l||c|c||c | c}
\hline
 & \multicolumn{2}{c||}{Change on}  \\
 %  \cline{1-5}\noalign{\smallskip}\\
phonemes & train data & test data \\
\hline
{\tt aa-ao} & 0.78 \%  & 1.82 \% \\
{\tt aa-dcl} & 0.09 \%  & 0 \% \\
{\tt aa-iy} & 0 \%  & 0 \% \\
{\tt aa-sh} & 0 \%  & 0 \% \\
{\tt ao-dcl} & 0.08 \%  & 0 \% \\
{\tt ao-iy} & 0 \%  & -0.17 \% \\
{\tt ao-sh} & 0 \%  & 0 \% \\
{\tt dcl-iy} & 2.48 \%  & 0.99 \% \\
{\tt dcl-sh} & 0 \%  & -0.48 \% \\
{\tt iy-sh} & 0.07 \%  & 0.19 \% \\
\hline
\end{tabular}
\caption{Improvement (positive) or worsening (negative) of performance of $B$-classifiers compared to $Z$-classifiers}
\label{tab:results1}
\end{center}
\end{table}

\begin{figure}[htb]
%\begin{minipage}[b]{0.48\linewidth}
%  \centering
%  \centerline{\includegraphics[width=8.5cm]{zero_ad.pdf}}
% % \centerline{(a) Result 1}\medskip
%\end{minipage}
%\begin{minipage}[b]{0.48\linewidth}
%  \centering
%  \centerline{\includegraphics[width=8.5cm]{zero_ai.pdf}}
%  %\centerline{(b) Result 1}\medskip
%\end{minipage}
%\begin{minipage}[b]{0.48\linewidth}
  \centering
  \centerline{\includegraphics[width=8.5cm]{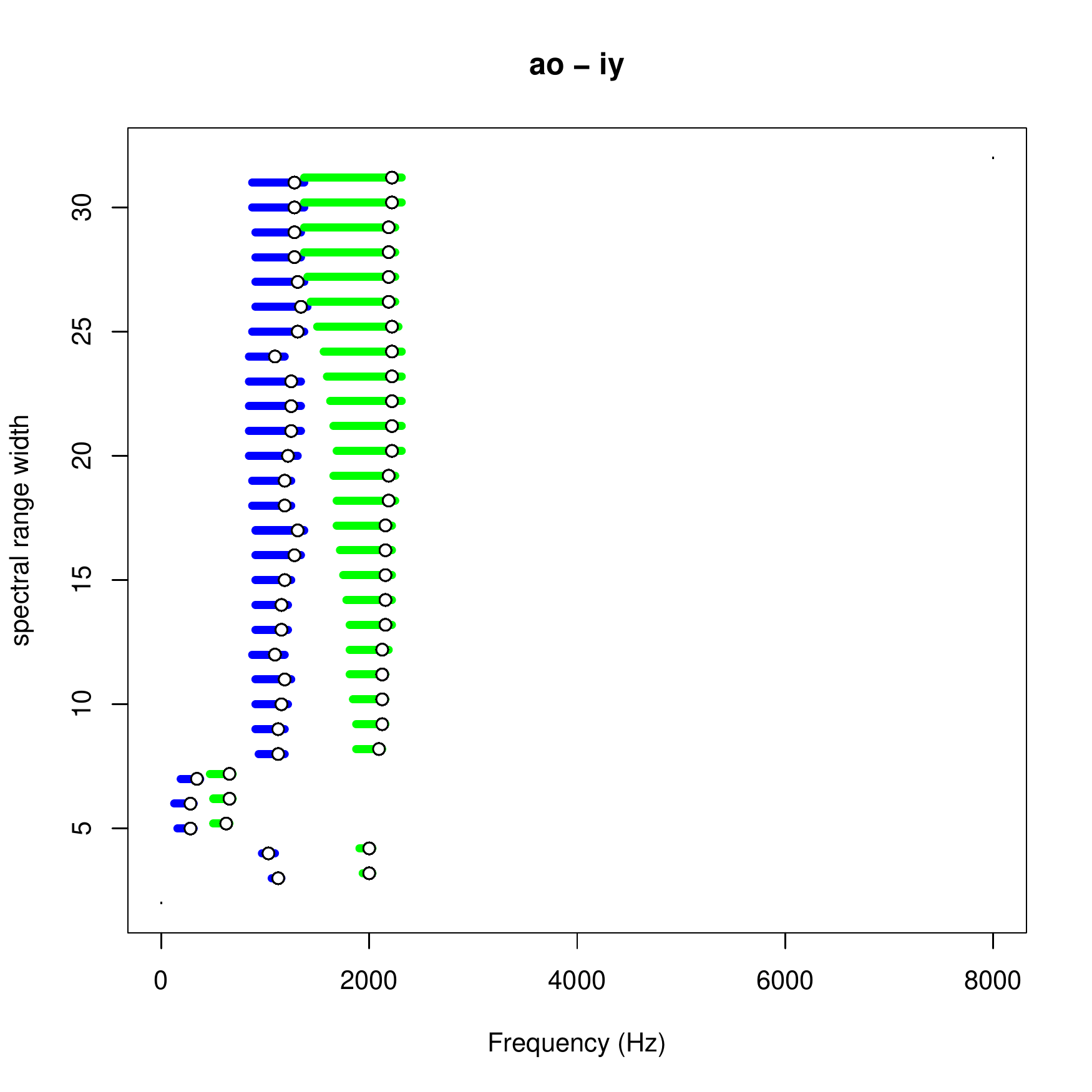}}
  %\centerline{(b) Result 1}\medskip
%\end{minipage}
%\begin{minipage}[b]{0.48\linewidth}
%  \centering
%  \centerline{\includegraphics[width=8.5cm]{zero_is.pdf}}
%  %\centerline{(b) Result 1}\medskip
%\end{minipage}
\vskip -0.3cm
\caption{Spectral ranges of optimal neural networks for discrimination {\tt ao-iy} plotted against increasing spectral range width. White circle indicates the position of quantile (right is the maximum, the left end and the middle would represent the minimum and the  median respectively). More graphs are available in \cite[page 83]{MinMaxBook}.}
\label{fig:graphs}
%\end{figure*}
\end{figure}

Unlike many other classes of neural networks, the structure (and not only response) of our networks can be clearly visualized as seen in Figure \ref{fig:graphs}.  In the figure in horizontal scale we indicate spectral ranges to which the two neurons are sensitive. In the vertical scale we indicate maximum length of spectral ranges $\vecr_1, \vecr_2$.

% \begin{figure*}[!htb]

\section{Continuous neighborhood of optimal quantile classifiers}

One may try to improve the discrimination results by allowing more general functions of a spectral range. So let us suppose that we have spectral ranges $\vecs_1$, $\vecs_2$. Let us write $sort(\cdot)$ for the function that orders its vector argument elementwise. Commonly used LDA tries to optimize Fisher's discriminant of functions 
\begin{align}
\vecw_1 \cdot \vecr_1 &- \vecw_2 \cdot \vecr_2,
\intertext{with real valued weight vectors $
\vecw_1, \vecw_2$, whereas in the previous section we optimized separations of expressions}
q_1(\vecr_1) - q_2(\vecr_2) &= \vecb_1 \cdot sort(\vecr_1) - \vecb_2 \cdot sort(\vecr_2)  \label{eq:b},
\end{align}
with exactly one nonzero entry in both $\vecb_1$ and  $\vecb_2$. One may relax this condition on $\vecb_i$ to obtain other classifiers.

\begin{example}
Consider the $B$-classifier defined by function 
\begin{align}
f(\vecs) = \max(s_{62}, s_{63}, \ldots, s_{74}) - s_1, \label{eq:diclassifier}
\intertext{with threshold value $\theta = 4.03279$ that decides}
\begin{cases}
\textrm{phoneme is 'dcl' if } f(\vecs) < 4.03279,\\
\textrm{phoneme is 'iy' if } f(\vecs) > 4.03279.
\end{cases}
\end{align}
We can consider vectors $\vecb_i$ in \eqref{eq:b} of the following categories
\begin{itemize}
\item (monotone OWA) nonnegative, increasing entries in each $\vecb_i$, with total sum equal to one,
\item (OWA \cite{YagerOWA}) nonnegative entries in each $\vecb_i$, with total sum equal to one,
\item (ordered LDA) arbitrary real valued entries.
\end{itemize}
Comparison of these methods  can be seen in Table \ref{tab:results2}. All methods in the table use only spectral compoments $s_1$ and the spectral range $s_{62}, s_{63}, \ldots, s_{72}$. Balanced LDA is a variant of $LDA$, in which the sum of all coefficients is 0. 
\begin{table}[!ht]
\begin{center}
\begin{tabular}{|l||p{1cm}|p{1cm}||p{1cm}|p{1.cm}|}
\hline
method & train error & test error & Fisher  train score & Fisher test score \\
\hline
quantiles ($f$)& 2.7 \%  & 5.1 \%  & 6.22 & 6.54\\
monotone OWA & 3 \%  & 4 \% & 6.24 & 6.5 \\
OWA & 3 \%  & 4.5 \%  & 6.28 & 6.54\\
ordered LDA & 0.8 \%  & 1.8 \% & 15 & 12.83 \\
\hline
\hline
balanced LDA & 4.2 \% & 5.7 \% &  5.69 & 5.85 \\
LDA \strut & 1.2 \% & 2.2 \% & 13.81 & 12.22\\
\hline
\end{tabular}
\caption{Comparison of LDA with ordered methods. See text for explanation of various methods.}
\label{tab:results2}
\end{center}
\end{table}
Our conclusion is that not much is gained by passing from discrete structures represented by quantiles to weighted ones, in line with similar research \cite{Soudry2013}.

%Comparison of numerical values of coefficients can be seen in Figure \ref{fig:weights}
%\begin{figure}[!htb]
%\begin{minipage}[b]{1.0\linewidth}
%  \centering
% \centerline{\includegraphics[width=8.5cm]{coeff}}
% \caption{Coefficients of various ordered methods}
% \label{fig:weights}
% \end{figure}

\end{example}
%%  \vspace{2.0cm}
%  \centerline{(a) Result 1}\medskip
%\end{minipage}

% pdf("Desktop/Memristor/ICASSP2014/coeff.pdf")
% xyplot(xm + x + lda ~ coeff, data = ord_weights, type = "o", key = simpleKey( c("monotone OWA", "OWA", "ordered LDA"), corner = c(0.2,0.2)), ylab="weight", xlab="coefficient")
% dev.off()

\section{Future work}

In this contribution we have opted to present only the simplest results due to limited space.
It is clear however that more research is needed for this class of networks to find applications. Let us outline the directions further research may take.

First, one should take into account known psychoacoustic phenomena of human hearing. It may prove advantageous to adjust spectral power to reflect varying sensitivity to varying frequency \cite{Morgan2012}. Our experiments showed that often low frequency power was crucial for discrimination and thus it may prove useful to use Q-transform \cite{BrownQTransform} which provides more data points in lower frequencies compared to ordinary FFT.

Secondly, it is well known that it is two and sometimes up to 4 formants that characterize a vowel. It is therefore necessary to consider a more complex set of discrimination functions. In \cite{KSAlgebraI} we proposed an algebra, whose elements are candidates for describing the structure of more complex networks.

Let us conclude with summarizing advantages of proposed networks. By design, they provide a guaranteed performance on variations of  trained data unseen during training, they are interpretable and have very low Kolmogoroff's complexity, as the example \eqref{eq:diclassifier} shows. 

\begin{figure}[!htb]
%\begin{minipage}[b]{1.0\linewidth}
  \centering
 \centerline{\includegraphics[width=8.5cm]{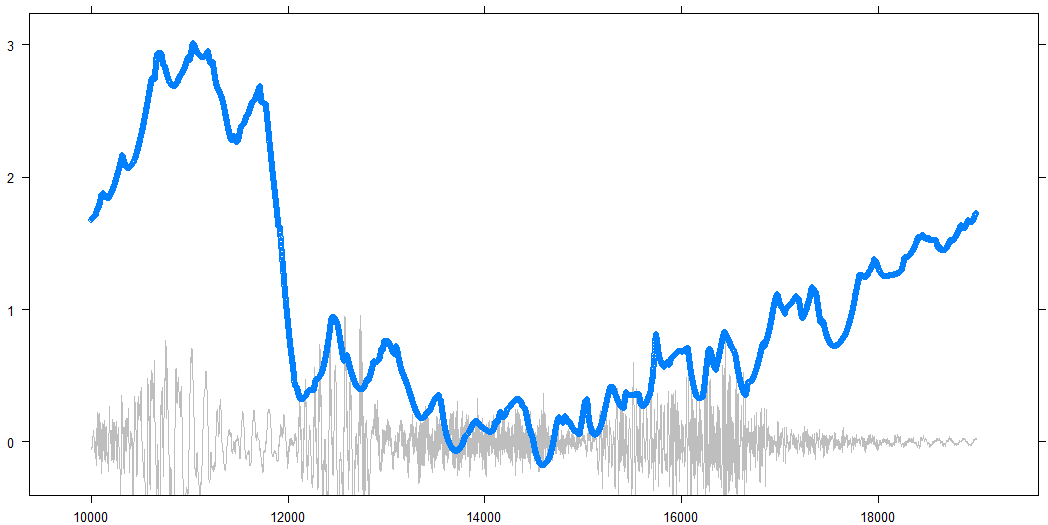}}
 \caption{Evaluation of `dcl-iy' classifier over a Slovak word (IPA: /\textipa{odiSla}/) superimposed on PCM signal. Note that the classifier was train on English data  (TIMIT).
}
 \label{fig:prechod}
 \end{figure}
Last, but not the least, the networks can be easily implemented in hardware, since BJT \cite{Yamakawa1}, CMOS \cite{Baturone97}  and even passive memristor implementations \cite{arxiv} of min, max and comparison operators exist. In this context it is worthwhile to point out that it is the change of, rather than the absolute spectral content, that can be read off from these discriminants. This is illustrated in Figure \ref{fig:prechod} where transition between phonemes is quite strong. One may thus hypothesize that analog hardware speech recognizer could be based on silicon cochlea (\cite{SiliconCochlea}, followed by processing by a neural network of the kind described here, whose output would be fed to adaptive differentiator like that of Delbr\"uck and Mead \cite{AdaptiveCircuit}, and finally to a memristive switch \cite{TheMissingMemristorFound}.

\printbibliography
\end{document}